# THE US MUON ACCELERATOR PROGRAM*

M.A. Palmer[#], Fermilab, Batavia, IL 60510, USA
*for the MAP Collaboration*

## Abstract

A directed R&D program is presently underway in the U.S. to evaluate the designs and technologies required to provide muon-based high energy physics (HEP) accelerator capabilities. Such capabilities have the potential to provide unique physics reach for the HEP community. An overview of the status of the designs for the neutrino factory and muon collider applications is provided. Recent progress in the technology R&D program is summarized.

## INTRODUCTION

The U.S. Muon Accelerator Program (MAP) was approved in 2011 as part of the response, by the Office of High Energy Physics of the U.S. Department of Energy, to recommendations from the 2008 U.S. Particle Physics Project Prioritization Panel (P5) [1]. The aim of the MAP effort is to develop the concepts required for the application of muon accelerators for high energy physics science needs and to assess the feasibility of the technologies required to support these applications. A key feature of muon accelerator capabilities is that a facility based on muon beams has the ability to support world-class research on both the Intensity and Energy Frontiers. The capabilities being developed by MAP range from those that are ready for immediate deployment, i.e., a short-baseline neutrino facility such as nuSTORM [2,3], to much longer timescale facilities such as a long-baseline neutrino factory or TeV-scale muon collider.

Over the course of the past 2 years, MAP has developed a detailed technical plan for moving forward with its program to evaluate the feasibility of muon accelerators for high energy physics applications. There are 3 principal elements of this plan:

- A *Muon Accelerator Staging Study (MASS)* working group was created in 2012 and charged with identifying a clear path towards future muon accelerator capabilities which would allow cost and technical risks to be controlled. The working group was also charged with reviewing each accelerator sub-system and to prepare a set of recommendations on specific design thrusts that should be pursued in order to optimize technical performance, risk and/or cost.
- An Initial Baseline Selection (IBS) process has been implemented to carry out the detailed development and evaluation of the design concepts required for each stage of a muon accelerator facility, from the proton driver through the neutrino factory and collider systems. The MASS recommendations are actively being incorporated into this process. The IBS process outlines a roughly 2.5 year effort to evaluate each of the major design concepts for a muon-based facility and to document those concepts at a level where they could serve as inputs to a future conceptual design for a muon accelerator facility.
- An R&D program to demonstrate critical concepts for muon accelerators and to validate the technologies critical to muon accelerator capabilities. A particular focus of this program is the development of the technologies required for a muon ionization cooling channel – an accelerator system unique to this type of facility. It also supports the initial demonstration of the ionization cooling process with the Muon Ionization Cooling Experiment (MICE) at the Rutherford Appleton Laboratory.

These three efforts are implemented in such a way that they can support a 2-phase feasibility assessment for high intensity muon accelerator capabilities. The first phase, 2013-2016, will establish the baseline design concepts for a muon-based facility and, along with the technology R&D effort, will provide the necessary specifications for a small set of critical technology demonstrations to take place during a second phase, 2017-2020. Thus, the MAP effort aims for a clear statement about the feasibility of muon accelerator concepts by the end of the decade. A clear determination of muon accelerator feasibility on this timescale would provide timely input for a decision by the high energy physics community on the need for a high precision neutrino source to support exploration of new physics in the neutrino sector and/or a muon collider to explore potential new physics at the several TeV energy scale.

## THE MUON ACCELERATOR STAGING STUDY (MASS)

The MASS working group has prepared a staged plan for muon accelerator capabilities that can support cutting edge science on both the Intensity and Energy Frontiers [4,5]. The potential facilities along this staged path are:

- nuSTORM: a short-baseline neutrino factory which could provide a definitive measurement of sterile neutrinos as well precision measurements of neutrino cross-sections in support of a long-baseline neutrino oscillation program;
- NuMAX: a long-baseline neutrino factory optimized for the 1300 km baseline from Fermilab to Sanford Underground Research Facility (SURF), which could initially operate with a 1 MW proton driver and no muon cooling;

___________________________________________
* Work supported by US DOE under contract DE-AC02-07CH11359.
#mapalmer@fnal.gov

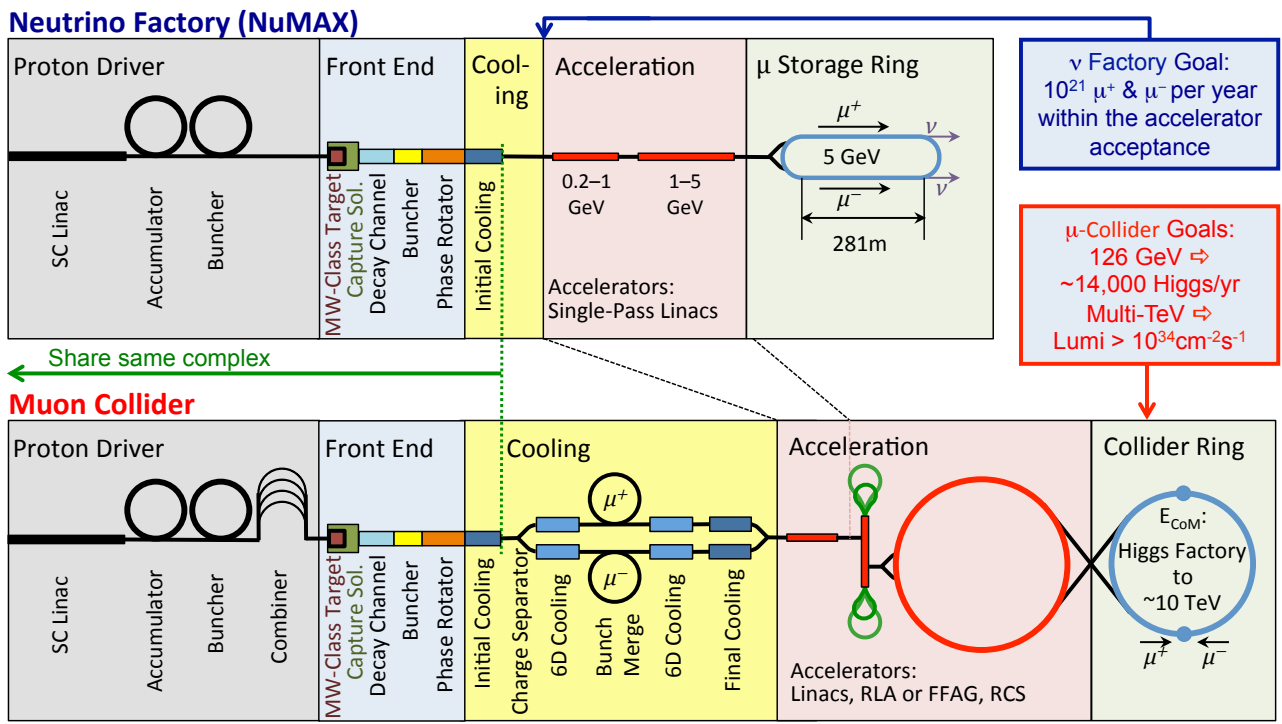

Figure 1: A block diagram showing the key systems needed for a long-baseline neutrino factory capability and a muon collider capability. Much of the infrastructure for each capability could be shared, thus enabling a cost effective multipurpose facility.

- NuMAX+: a full intensity neutrino factory with additional proton beam power and 6D muon cooling as an upgrade to NuMAX. This facility would offer the ultimate precision microscope for the study of CP violation and possible new physics in the neutrino sector;
- Higgs Factory: a collider which, in its baseline configuration, is capable of producing roughly 13,500 Higgs particles per Snowmass year ($10^7$ s) with exquisite energy resolution;
- A multi-TeV Muon Collider: a machine which could be operated in the 1-10 TeV center-of-mass energy range to carry out precision exploration of new physics that may be observed at the LHC. Above roughly 1.5 TeV, such a machine is expected to offer the best performance to cost ratio for providing lepton collider capabilities for the HEP community.

The deployment of a facility with some or all of these stages enables a scenario where the systems and performance evaluations for each subsequent stage can be supported by R&D based on the beams provided by the already existing stages.

## THE INITIAL BASELINE SELECTION (IBS) PROCESS

The MAP IBS process is intended to deliver clearly defined concepts for each major system required for a muon accelerator facility. Its scope spans all of the systems required for short- and long-baseline neutrino factories as well as collider options. Figure 1 shows the block diagram of the systems required to support a long-baseline neutrino factory as well as those necessary for a muon collider. As is clear from the diagram, significant synergies exist between the two. Thus the IBS process aims to develop concepts that are compatible with a multi-purpose facility. As already noted, the MASS working group has provided a number of recommendations for design thrusts to best enable this multi-purpose configuration and which also can minimize cost and risk in the ultimate designs. These recommendations form an integral of the IBS process.

If fully funded, it is anticipated that a complete set of baseline concepts for these capabilities can be specified on the 2016 timescale. The availability of these concepts would then set the stage to complete a set of technology demonstrations with clear technical specifications as part of the second phase of the MAP Feasibility Assessment. In combination with the results from the MAP R&D efforts, including the MICE demonstration of ionization cooling with RF re-acceleration in the last half of the decade, this would enable a clear statement on muon accelerator feasibility for HEP applications by 2020.

## THE MAP R&D EFFORT

MAP R&D efforts fall into two principal categories. The first is technology R&D that spans the development of: normal conducting RF cavities capable of operating in high (multi-Tesla) magnetic fields; superconducting RF cavities suitable for use in an ulta-fast muon acceleration chain; very high field magnets including those utilizing

high temperature superconductors; rapid-cycling magnets for the ultra-fast muon acceleration chain; and high power target and absorber concepts. The research into RF cavities operating in magnetic fields is supported by the MuCool Test Area (MTA) experimental facility at Fermilab. The second major category of R&D is large-scale system and physics demonstrations to validate key concepts. The two principal efforts of this type have been the MERIT demonstration of liquid metal jet technology for targets capable of handling multi-MW incident beam power [6] and the MICE experiment [7] aimed at the explicit demonstration of ionization cooling of muons. At the conclusion of the IBS process, the MAP R&D plan envisions a major system demonstration of the technology required for a high performance 6D ionization cooling channel.

Over the past two years, significant progress has been achieved in the operation of RF cavities in magnetic fields. Recent successes in operating vacuum RF cavities in this environment [8] and preparations to test a new cavity incorporating key new design elements for operation in high field [9] have been presented. Updates on the use of high-pressure gas-filled cavities have also been presented [10,11,12]. In both cases, the results show very significant progress towards RF designs capable of meeting the neutrino factory and muon collider performance requirements. These results are being incorporated into modern 6D cooling channel designs [13,14].

Another area of significant progress has been in preparations for executing the MICE experiment. MICE construction activities are moving forward rapidly and next year will see the start of a major experimental run [15,16].

## RECENT EVENTS

Although the MAP R&D plan was recently endorsed at a major review held by the US DOE in February 2014, the new P5 report in the US has recommended ramping down muon accelerator activities in the US accelerator R&D portfolio [17]. In this context, the MAP collaboration is working to prepare a revised proposal to enable the most critical demonstrations to be completed and the key concepts developed over the last few years to be fully documented so that a viable effort could be restarted in the future based on the physics needs. Unfortunately, this approach risks the loss of critical expertise in this novel area of accelerator physics.

## CONCLUSION

Significant progress towards the demonstration of the concepts and technologies required for muon accelerator systems has taken place over the last two years. A clear plan for completing the feasibility assessment for long-baseline neutrino factories and muon colliders by the end of the decade has been developed. However, recent developments in the US high energy physics program place this progress in jeopardy. It is anticipated that a revised plan for muon accelerator R&D efforts will become clear by the end of 2014.